
\magnification = 1200
\hsize = 15 truecm
\vsize = 23 truecm
\baselineskip 20 truept
\voffset = -0.5 truecm
\parindent = 1 cm
\overfullrule = 0pt
\count0 = 1
\footline={\hfil}

\null

\settabs 3 \columns
\+&&  Preprint-KUL-TF-93/47 \cr
\+&&  hep-th/9311054 \cr
\+&&  November 1993 \cr

\null
\vskip 1 truecm

\centerline
{\bf  Topological gravity with exchange algebra}

\vskip 1 truecm

\centerline{\bf S. Aoyama}
\smallskip

\vskip 0.5 cm

\centerline{\bf Instituut voor Theoretische Fysica}
\smallskip
\centerline{\bf Katholieke Universiteit Leuven}
\smallskip
\centerline{\bf Celestijnenlaan 200D}
\smallskip
\centerline{\bf B-3001 Leuven, Belgium}

\vskip 2 truecm

\noindent
{\bf Abstract}

A topological gravity is obtained by twisting the  effective $(2,0)$
super\-gravity. We show that this topological gravity has an infinite number of
BRST invariant quantities with conformal weight $0$. They are a tower of
OSp$(2,2)$ multiplets and satisfy the classical exchange algebra of OSp$(2,2)$.
We argue that these BRST invariant quantities become physical operators in the
quantum theory and their correlation functions are braided according to the
quantum OSp$(2,2)$ group. These properties of the topological effective gravity
are not shared by the standard topological gravity.

\vskip 1.5 cm
\noindent
e-mail: shogo\%tf\%fys@cc3.kuleuven.ac.be

\vfill\eject

\footline={\hss\tenrm\folio\hss}

It is one of the important subjects in the string theory to understand coupled
systems of $2$-dimensional gravity and conformal field theories (CFTs). There
is much interest in the case where they are topological, i.e., topological
gravity coupled with topological conformal field theories (TCFTs), since the
analysis is simplified by the BRST symmetry and  becomes
far-reaching$^{[1,2]}$.
Topological gravity was formulated in refs 3 and 4. In this note we give one
more different formulation to discover some new properties. Namely we think of
$N=2$ supergravity coupled with a $N=2$ CFT. The former gets dynamical through
quantization, i.e., effective $N=2$ supergravity. By twisting$^{[1,5]}$ at this
stage the total system turns into a topological effective gravity coupled with
a TCFT.
 The novelty of this topological gravity is that there exist an infinite number
 of BRST invariant quantities with conformal weight $0$. They are a tower of
OSp$(2,2)$ multiplets and satisfy the classical exchange algebra of OSp$(2,2)$.
It is natural to assume that in the quantum theory these BRST invariant
quantities  become physical operators. Then they define the physical ground
states which
are  an infinite tower of OSp$(2,2)$ multiplets (, precisely speaking, quantum
OSp$(2,2)$ multiplets). Moreover we may argue that correlation functions of
these physical operators are braided according to the quantum OSp$(2,2)$ group.
These ground states  are purely gravitational. They contrast with the ground
ring of the $c=1$ string$^{[6]}$. They are rather gravitational descendants of
the puncture operator$^{[1,2]}$ in the language of the standard topological
gravity.

The untwisted effective $N=2$ supergravity can be   discussed in one of the
following formulations
\item{i)} light-cone gauge formulation$^{[7]}$;
\item{ii)} geometrical formulation$^{[8]}$;
\item{iii)} formulation by the constrained WZWN model$^{[9]}$;
\item{iv)} conformal gauge formulation$^{[10,11]}$.

\noindent
In ref. 12 the non-supersymmetric effective gravity was discussed  in all these
formulations. The consistency among them was checked at both classical and
quantum levels. A similar consistency check would be done in the $N=1$ or $N=2$
case as well. In this note we choose the geometrical formulation. We also
choose to study the effective $(2,0)$ supergravity for simplicity. By the
twisting procedure this becomes a topological gravity, which is topological  in
the supersymmetric sector. We discuss that in this sector the theory is indeed
such as described just above.

The supersymmetric sector of the effective $(2,0)$ supergravity has
characteristic properties in the geometrical formulation, namely the
superconformal symmetry and the exchange algebra of OSp$(2,2)$. They are
essential for our arguments. The theory  also exhibits remarkable properties in
the non-supersymmetric sector like the Knizhnik-Zamolodchikov equation (,
simply the KZ equation,) and the  Kac-Moody symmetry with OSp$(2,2)$.
But this sector is irrevalent for our purpose in this note. So these issues
will not be discussed. Suffice it to say that  the appearances of the KZ
equation and the Kac-Moody symmetry with OSp$(2,2)$ are almost evident by
taking analogy to the non-supersymmetric effective gravity in the geometrical
formulation$^{[12]}$. Similar things have been originally shown  by studying
the non-supersymmetric sector of the effective $(0,0)$ and $(1,0)$
supergravities in the light-cone gauge$^{[7]}$. But we would like to stress
that the light-cone gauge approach does not give any proper account for the
supersymmetric sector as the geometrical formulation does$^{[12]}$. This is the
reason why we choose the latter formulation in this note.

\vskip 1cm

Let us begin by summarizing the $(2,0)$ superconformal group$^{[13]}$. The
supersymmetric sector of the (2,0) superspace is described by a real
coordinate $x$, a fermionic complex coordinate $\theta^+$ and its
complex conjugate $\theta^-$, while the non-supersymmetric sector by a
real coordinate $t$ alone. We define the superderivatives as
$$
D_+ = {\partial \over \partial \theta^+}  +
\theta^-\partial_x,\quad\quad
D_- = {\partial \over \partial \theta^-}  + \theta^+\partial_x,
$$
which satisfy
$$
\{D_+,D_-\} = 2\partial_x,  \quad \quad       \{D_{\pm},D_{\pm}\} = 0.
$$
An element of the $(2,0)$ superconformal group is given by superdiffeomorphisms
$$
\eqalign{x  \longrightarrow f(x,\theta^+,& \theta^-;t),  \quad \quad \quad
\theta^{\pm} \longrightarrow \varphi^{\pm}(x,\theta^+,\theta^-;t), \cr
t & \longrightarrow g(x,\theta^+,\theta^-;t), \cr} \eqno (1)
$$
which change the derivatives in the respective sector as
$$
D_{\pm}  = (D_{\pm}\varphi^+)D_+^\varphi \ + \ (D_{\pm}\varphi^-)D_-^\varphi ,
\eqno (2)
$$
$$
\partial_t  = (\partial_t g)\partial_g,
$$
with
$$
D_+^\varphi = {\partial \over \partial \varphi^-} + \varphi^-{\partial
\over \partial f}, \quad \quad
D_-^\varphi = {\partial \over \partial \varphi^+} + \varphi^+{\partial
\over \partial f},
$$
$$
\partial_g = {\partial \over \partial g }.
$$
This imposes the following conditions on superdiffeomorphisms
$$
\eqalignno{D_{\pm}f & = \varphi^+D_{\pm}\varphi^- + \varphi^-D_{\pm}\varphi^+.
& (3) \cr
\partial_x g & = \partial_{\pm} g = 0.  & (4)  \cr}
$$
Since $D_{\pm}^2 = 0$, the superconformal
conditions (3) imply that either $D_{\pm}\varphi^{\pm} = 0$ or
$D_{\pm}\varphi^{\mp} = 0$. For convenience we will choose the case
$$
D_{\pm} \varphi^{\mp} = 0,  \eqno (5)
$$
in this note. Then eqs (2) and (3) are reduced respectively to
$$
D_+ = (D_+\varphi^+) D_+^\varphi,  \quad \quad \quad    D_- = (D_-\varphi^-)
D_-^\varphi,
$$
and
$$
D_+f  = \varphi^-D_+\varphi^+, \quad \quad \quad  D_-f = \varphi^+D_-\varphi^-
.  \eqno (6)
$$
In terms of infinitesimal diffeomorphisms $\delta f$ and $\delta \varphi^{\pm}$
the superconformal conditions (6) are written in the forms
$$
\eqalign{\delta \varphi^+ & = {1\over D_-\varphi^-}D_-(\delta f +
\varphi^+\delta \varphi^-),  \cr
\delta \varphi^- & = {1\over D_+\varphi^+}D_+(\delta f + \varphi^-\delta
\varphi^+). \cr}  \eqno (7)
$$
The $(2,0)$ superconformal  transformations are obtained as solutions of the
equations (4) and (7):
$$
\eqalignno{
\delta f & = [\xi\partial_x + {1\over2}(D_+\xi) D_- + {1\over2}(D_-\xi) D_+]
f\quad  + \quad  \zeta\partial_t f,  & (8) \cr
\delta \varphi^{\pm} & = [ \xi\partial_x + {1\over2}(D_{\mp}\xi)
D_{\pm}]\varphi^{\pm}\quad + \quad \zeta\partial_t \varphi^{\pm}, & (9)  \cr
\delta g & = \zeta \partial_t g. &  \cr}
$$
Here $\xi$ and $\zeta$ are infinitesimal  parameters of the $(2,0)$
superconformal transformations in the super- and non-supersymmetric sectors
such as $\xi (x,\theta^+,\theta^-)$ and $\zeta (t)$. Note that $\xi$ is given
by
$$
\xi = \delta x + \theta^+ \delta \theta^- + \theta^- \delta \theta^+, \quad
\quad \zeta = \delta t,   \eqno (10)
$$
which follows from eqs (8) and (9).
A $(2,0)$ superconformal field with weight $(h,$\ $w)$ is defined as
transforming by (1) according to
$$
\Psi (x,\theta^+,\theta^-,t) = \Psi (f,\phi^+,\phi^-, g) (D_+\varphi^+)^{h-{q
\over 2}} (D_-\varphi^-)^{h+{q \over 2}} (\partial_t g)^w ,
$$
in which $q$ is $U(1)$ charge in the supersymmetric sector. It becomes
infinitesimally
$$
\eqalign{\delta \Psi & =  \  \{\xi \partial_x + {1 \over 2}(D_+\xi) D_- + {1
\over 2}(D_-\xi) D_+ + (h \partial_x \xi + {q \over 4}[D_-,D_+]\xi)  \cr
& \ + \ \zeta \partial_t \ + \ w(\partial_t \zeta)\  \} \Psi,   \cr}    \eqno
(11)
$$
by using (10).

\vskip 1truecm

In the geometrical formulation the effective (2,0) supergravity is formulated
as a $2$-dimensional field theory on the coadjoint orbit of the $(2,0)$
superconformal group$^{[14]}$. The action is given by
$$
S = {k \over 2\pi} \int dxdtd\theta^+d\theta^-\ \partial_t(\log
D_+\varphi^+)\log
D_-\varphi^- ,   \eqno (12)
$$
in which $\varphi^{\pm}$ are fermionic superfields characterized above.
By the construction this action is invariant under the (2,0) superconformal
transformations (9). In fact we find that

$$
\delta S = - {1 \over 2\pi} \int dxdtd\theta^+d\theta^- [
\xi(x,\theta^+,\theta^-)\partial_t {\cal S} + \zeta(t)\partial_x {\cal T}] = 0,
$$

\noindent
where

$$
\eqalign{
{\cal S }  = & k[\partial_x(\log D_+\varphi^+ - \log D_-\varphi^-) +
2{\partial_x\varphi^+ \over D_+\varphi^+}{\partial_x\varphi^- \over
D_-\varphi^-}] \cr
& (\rm{N=2\ super\  Schwarzian\ derivative}), \cr}  \eqno (13)
$$
\noindent
and
$$
{\cal T }  = 2k{\partial_t\varphi^+ \over D_+\varphi^+}
{\partial_t\varphi^- \over D_-\varphi^-}.   \eqno (14)
$$

\noindent
In this note we concentrate on the supersymmetric sector which exhibits a
topological nature later on. In ref. 15 the Poisson brackets of $\varphi^{\pm}$
and $f$ with themselves were worked out in this sector, i.e., on a plane at a
given time $t$.
They were found as solutions to the following requirements:
\item{i)} the super Schwarzian derivative (13) is the generator of the
superconformal transformations given by (8) and (9);
\item{ii)} the Poisson brackets are consistent with the superconformal
conditions (6);
\item{iii)} the Jacobi identities are satisfied.

\noindent
It turned out that
$$
\eqalign{
\{\varphi^+(z),\varphi^-(z')\}_{PB}  = -{\pi \over 2k}[\ &
\theta(x-x')\{\ f(z)\ +\
\varphi^+(z)(\ \varphi^-(z) - \varphi^-(z')\ )\ \}  \cr
+ &  \theta(x'-x)\{\ f(z')\ +\ \varphi^-(z') (\ \varphi^+(z') -
\varphi^+(z)\ )\ \}\ ],\cr}
$$

$$
\eqalign{
\{f(z),\varphi^+(z')\}_{PB} = -{\pi \over 2k}[\
\theta(x-x') & f(z)(\ \varphi^+(z) - \varphi^+(z')\ ) \cr
+  \theta(x'-x) & \{\ f(z)\varphi^+(z')\ +\
f(z')(\ \varphi^+(z) - 2\varphi^+(z')\ )   \cr
+ & \  \varphi^+(z)\varphi^+(z')\varphi^-(z')\ \}\ ], \cr}
$$

$$
\eqalign{
\{f(z), \varphi^-(z')\}_{PB} = -{\pi \over 2k}[
\theta(x-x') & f(z)(\ \varphi^-(z) - \varphi^-(z')\ )  \cr
+  \theta(x'-x) & \{\ f(z)\varphi^-(z')\ +\
f(z')(\ \varphi^-(z) - 2\varphi^-(z')\  )    \cr
+ & \  \varphi^-(z)\varphi^-(z')\varphi^+(z')\ \}\ ],  \cr}   \eqno (15)
$$

$$
\eqalign{
\{f(z),f(z')\}_{PB} = -{\pi \over 2k}[\ \theta(x-x')\{ & 2f(z)(\ f(z)
-  f(z')\ ) \cr
& - f(z)(\ \varphi^+(z)\varphi^-(z') +
\varphi^-(z)\varphi^+(z')\ )\ \}    \cr
+ \theta(x'-x)\{\  & 2f(z')(\ f(z)\ -\ f(z')\ ) \cr
& - f(z')(\ \varphi^+(z)\varphi^-(z') +
\varphi^-(z)\varphi^+(z')\ )\ \}\ ]  .    \cr}
$$

\vskip 0.5cm

\noindent
in which $z = (x,\theta^+,\theta^-)$ and
the t-dependence of $\varphi^{\pm}(z;t)$ and $f(z;t)$ has not
explicitly be written down.
We have normalized the coupling constant $k$ by requiring that the last Poisson
bracket tends to the one of the $(0,0)$ theory at the non-supersymmetric
limit$^{[12]}$.
By using these Poisson brackets we find the superconformal transformation of
the super Schwarzian derivative itself:
$$
\eqalign{  {1 \over 2\pi} \int  d z' & \xi (z')  \{{\cal S}(z'), {\cal S}(z)
\}_{PB}  \cr
& = [\ \xi (z) \partial_x + {1 \over 2}(D_+\xi (z)) D_- + {1 \over 2}(D_-\xi
(z)) D_+ + \partial_x \xi (z) \ ] {\cal S}(z) \cr
& + { k \over 2 } [D_+,D_-] \partial_x \xi (z), \cr}
$$
which is anomalous by the last piece. It is typical in the geometrical
formulation that the superconformal anomaly  already appears by a classical
calculation.

\vskip 1cm

The action (12) has a hidden symmetry in the non-supersymmetric sector.
Consider the non-linear OSp$(2,2)$ transformations given by$$
\eqalign{
\delta f = & \epsilon^1 \ + \epsilon^0f \ + \epsilon^{-1}f^2 \cr
& + \varphi^+(\epsilon^{1\over 2}_+\ + \epsilon^{-{1\over 2}}_+ f)
+ \varphi^-(\epsilon^{1\over 2}_- \ + \epsilon^{-{1\over 2}}_-f), \cr}
\eqno (16)
$$
with infinitesimal parameters $\epsilon^i = \epsilon^i(t)$, $i =
0,\pm{1\over 2}, \pm 1$. Then the transformation laws of the variables
$\varphi^{\pm}(z,t)$ may be found with recourse to the superconformal
conditions (6)$^{[15]}$:
$$
\eqalign{
\delta \varphi^+ & = {1\over 2}\epsilon^0\varphi^+ +
\epsilon^{-1}f\varphi^+ + \epsilon^{1\over 2}_-
+ \epsilon^{-{1\over 2}}_-(f + \varphi^+\varphi^-) +  {1\over
2}\epsilon\varphi^+,  \cr
\delta \varphi^- & = {1\over 2}\epsilon^0\varphi^- +
\epsilon^{-1}f\varphi^- + \epsilon^{1\over 2}_+
+ \epsilon^{-{1\over 2}}_+(f - \varphi^+\varphi^-) -  {1\over
2}\epsilon\varphi^-,  \cr}   \eqno (17)
$$
in which $\epsilon$ is a $U(1)$ parameter of the OSp$(2,2)$ transformations
such as $\epsilon = \epsilon (t) = -\overline \epsilon$. We can show that the
action (12) is invariant by (17)$^{[15]}$.
As for the $(0,0)$ effective gravity$^{[12]}$, this local symmetry  may be
similarly exploited to show the KZ equation and the Sugawara form of the
energy-momentum tensor (14) with OSp$(2,2)$ in the non-supersymmetric sector.
But we are not involved in the issue in this note. Therefore hereinafter it is
treated simply as a global symmetry in the supersymmetric sector.

\vskip 1cm

In ref. 15 it was shown that the quantities
$$
\psi =
\left( \matrix{ \psi_{\ \ { 1\over 2}} \cr
                \psi_{-{1\over 2}} \cr
                \psi_{\ \ 0}  \cr} \right) =
{1\over D_+\varphi^+} \left(
\matrix{f - \varphi^+\varphi^- \cr
1  \cr
\varphi^-  \cr}
\right) ,  \quad\quad  c.c.,      \eqno (18)
$$
have remarkable properties. First of all they
form the lowest dimensional representation of OSp$(2,2)$. Namely by the
transformations (16) and (17) we find
$$
\delta \psi = [\sum_{i=0,\pm 1}{\epsilon^iL_i} + \sum_{i=\pm{1 \over
2}}{(\epsilon^i_+L^+_i + \epsilon^i_-L^-_i)} + \epsilon L\  ]\psi.   \eqno (19)
$$
\noindent
Here $L_i\ (i=0,\pm 1), L^{\pm}_i\ (i=\pm {1\over 2})$ and $L$ are the
generators of OSp$(2,2)$ given by $3\times 3$ matrices. The reader may refer to
ref. 15 for the explicit expressions.
Secondly the quantities (18) obey the classical exchange algebra:
$$
\{\psi(z)^{\otimes}_, \psi(z')\}_{PB} = {\pi \over k}[\ r^+\theta(x-x') +
r^-\theta(x'-x)\ ]\psi(z)\otimes \psi(z'),  \eqno (20)
$$

\noindent
by means of the Poisson brackets (15),
in which the $r$-matrices are given by
$$
r^+  =\  L_0\otimes L_0\ -\ L_{-1}\otimes L_1\ -\ {1\over 2}L^-_{-{1\over
2}}\otimes L^+_{1\over 2}\ - \  {1\over 2}L^+_{-{1\over 2}}\otimes L^-_{1\over
2}\ -\  L\otimes L,
$$
$$
r^-  = -L_0\otimes L_0\ +\  L_1\otimes L_{-1}\ -\ {1\over 2}L^+_{1\over
2}\otimes L^-_{-{1\over 2}}\ -\ {1\over 2}L^-_{1\over 2}\otimes
L^+_{-{1\over 2}}\ +\ L\otimes L,
$$

\noindent
and satisfy the classical Yang-Baxter equation
$$
[r_{12},r_{13}\} + [r_{12},r_{23}\} + [r_{13},r_{23}\} = 0.
$$
(There are misprints in the forefactor of the classical exchange algebra and
the fourth piece of $r^{\pm}$ given in ref. 15.)
Poisson brackets, a transformation law and an exchange algebra, such as given
by (15), (19) and (20) respectively, were firstly discussed for the
non-supersymmetric effective gravity in the geometrical formulation $^{[16]}$.
Then they were extended to the $(1,0)$ and $(2,0)$ supersymmetric
cases$^{[15,17]}$. But similar equations had been originally found for the
Liouville theory$^{[10,18]}$.

\vskip 1cm

We can make up higher dimensional representations of OSp$(2,2)$ out of the
lowest one $\psi$.
To this end let us introduce another set of superdiffeomorphisms $a$ and
$\beta^{\pm}$, which are constrained by
the same superconformal conditions as $f$ and $\varphi^{\pm}$.
Define a vector such that
$$
\overline \chi =
\left( \matrix{ \overline \chi_{\ \ { 1\over 2}} \cr
                \overline \chi_{-{1\over 2}} \cr
                \overline \chi_{\ \ 0}  \cr} \right) =
{1\over D_-\beta^-} \left(
\matrix{a - \beta^-\beta^+ \cr
1  \cr
\beta^+  \cr}
\right).
$$
It transforms as the complex conjugate of $\psi$:
$$
\delta \overline \chi = [\sum_{i=0,\pm 1}{\epsilon^iL_i} + \sum_{i=\pm{1 \over
2}}{(\epsilon^i_-L^+_i + \epsilon^i_+L^-_i)} - \epsilon L\  ]\overline \chi.
$$
The product
$$
\overline \chi \cdot \psi \equiv \overline \chi_{1 \over 2} \psi_{-{1 \over 2}}
-
\overline \chi_{-{1 \over 2}} \psi_{1 \over 2} + 2\overline \chi_0 \psi_0
$$
is  OSp$(2,2)$ invariant. By expanding the multiple product
$(\overline \chi \cdot \psi)^{2j}$,\ $j = {1\over 2}$ $,1,{3 \over 2}, \cdots$,
we find

$$
\eqalign{(\overline \chi \cdot \psi)^{2j} & \equiv \overline \chi^j \cdot
\psi^j  \cr
& \equiv \sum_{m = -j}^j (-1)^{j+m} \overline \chi_{-m}^j \psi_m^j +
2 \sum_{\mu = -j + {1 \over 2}}^{j-{1 \over 2}} (-1)^{j+\mu -{1 \over 2}}
\overline \chi_{-\mu}^j \psi_\mu^j,   \cr}
$$
with
$$
\eqalignno{
\psi^j & =
\left( \matrix{ \psi_m^j  \cr
               \psi_\mu^j  \cr} \right)  &  \cr
& = { f^j  \over (D_+ \phi^+)^{2j} }  \left(
\matrix{ \sqrt {{(2j)! \over (j+m)!(j-m)!}}
[f^m - (j+m)f^{m-1}\varphi^+\varphi^- ]   \cr
 \cr
 \cr
\sqrt {{(2j)! \over (j+\mu-{1 \over 2})!(j-\mu -{1 \over 2})!}}
f^{\mu -{1 \over 2}}\varphi^-  \cr  }
\right),  & (21) \cr}
$$

\noindent
and a similar expression for $\chi^j$ in terms of $a$ and $\beta^{\pm}$.
It is obvious that
the quantities $\psi^j$ form the $(4j+1)$-dimensional multiplets of OSp$(2,2)$.
Examining $\delta \psi^j$ by  (16) and (17)  we find the OSp$(2,2)$ generators
as $(4j+1)$ $\times (4j+1)$ matrices.
The components of the multiplets (21) now satisfy the classical exchange
algebra (20) with the $r$-matrices in the $(4j+1)$-representation.

\vskip 1cm

The OSp$(2,2)$ multiplets $\psi^j$ have further important properties. First of
all they are chiral$^{[20]}$:
$$
D_+ \psi^j = 0.    \eqno (22)
$$
We examine the transformation property by the superconformal transformations
(8) and (9). It is easily shown that they are the $(2,0)$ superconformal fields
 with weight $(-j,0)$ and $U(1)$ charge $2j$ which transform according to eq.
(11). By expanding the super Schwarzian derivative (13) we find the generators
of the $N=2$ superconformal symmetry:
$$
{\cal S } = J \ + \  \theta^+ G_+ \ + \  \theta^- G_- \ + \
 \theta^+\theta^- T. \eqno (23)
$$
The global supersymmetry transformation, generated by the supercurrent $G_+$,
will be of particular importance for our arguments soon later. We quickly look
into the transformation property of the multiplets $\psi^j$ by this. The
supercurrent $G_+$ generates a translation in the superspace such that
$$
\delta \theta^+ = {\rm const.}, \quad \quad  \delta \theta^- = \delta x =
\delta t = 0.
$$
{}From eq. (10) we have $ \xi = \theta^- \delta \theta^+$. Then the
transformation law (11) is reduced to
$$
\delta \psi^j = \theta^-\delta \theta^+ \partial_x \psi^j.  \eqno (24)
$$
for $\psi^j$. Taking into account the chirality condition we expand $\psi^j$ as
$$
\psi^j = u^j  \ + \  \theta^- \rho^j \ - \  \theta^+ \theta^- \partial_x u^j.
$$
With this  the global supersymmetry transformation (24) reads in components
$$
\delta u^j = 0,  \quad \quad \delta \rho^j = -\delta \theta^+ \partial_x u^j.
\eqno (25)
$$

\vskip 1cm

Let us now twist$^{[1,5]}$ the theory to get a topological gravity. The
energy-momentum tensor $T$ in eq. (23) is modified by adding the $U(1)$ current
$J$:
$$
T \longrightarrow T \ + \ {1 \over 2} \partial_x J.
$$
Remarkably the multiplets
 $\psi^j$, given by eq. (21), all become chiral superconformal fields with
 weight $(0,0)$ with respect to this modified energy-momentum tensor. The
supersymmetry current $G_+$ turns into the BRST current. As the result the
global supersymmetry transformation (25) is identified with the BRST
transformations.
We turn to quantization of this topological theory. All the properties of
$\psi^j$ hitherto found are well-based on the geometric and algebraic
arguments. Therefore it is fairly natural to suppose that they are maintained
at the quantum level. Precisely speaking we assume that by quantization
 (i) the superconformal symmetry is took over, and (ii)
 the classical exchange algebra (20) becomes the operator relation
$$
\eqalign{\psi^{j_1}_{m_1} (x) \psi^{j_2}_{m_2} (y) & =
\theta (x-y)  (R^+_q)^{m'_1 m'_2}_{m_1 m_2}
\psi^{j_2}_{m'_2} (y) \psi^{j_1}_{m'_1} (x)   \cr
& +  \theta (y-x)(R^-_q)^{m'_1 m'_2}_{m_1 m_2}
\psi^{j_2}_{m'_2} (y) \psi^{j_1}_{m'_1} (x), \cr}  \eqno (26)
$$
in which $R^{\pm}_q $ are the $R$-matrices of the quantum
OSp$(2,2)$ group with the deformation parameter
$$
q = \exp ({ \pi i \over k-1 }).
$$
Here the coupling constant $k$ has been shifted by the quadratic Casimir of the
adjoint representation of OSp$(2,2)$. It is a quantum effect which may be found
through analyses of the opposite sector$^{[12]}$. Keep in mind that in this
note as well as in ref. 12 the sign of the coupling constant $k$ is chosen to
be opposite to that of refs 7. (If the multiplets have different dimensions
$j_1$ and $j_2$, the $R$-matrices should be represented in a dimention which
equals a common multiple of $j_1$ and $j_2$.)
 The relation (26) correctly reproduces eq. (20) at the classical limit $k
\rightarrow \infty$, since we have
$$
R_q^{\pm} = 1 + {\pi i \over k-1 } r^{\pm} + O({1 \over (k-1)^2 }).
$$
It is a barely possible quantum generalization of the classical algebra (20)
which is consistent with the OSp\  $(2,2)$ and superconformal symmetries.
 It was successful to quantize the non-supersymmetric effective gravity  this
way$^{[19,12]}$.  We shall see the consequences of these assumptions for the
topological gravity.
 By  the assumption (i) the multiplets $\psi^j_m$  are chiral primaries of the
$N=2$ superconformal algebra at the quantum level. We have classically shown
that their conformal weight is zero after twisting. In principle
 it would be shifted by $\Delta$ due to quantum effects. As it was discussed in
refs 19 and 21, the quantum group (assumption (ii)) and conformal symmetries
(assumption (i)) interplay to fix this anomalous dimension:
$$
\Delta = -{c_j \over 2(k-1)},
$$
in which $c_j$ is the quadratic Casimir of the $(4j+1)$-dimensional
representation. Note that the anomalous dimension $\Delta$ tends to zero at the
classical limit $k \rightarrow \infty$.
A speciality about OSp$(2,2)$ is that the quadratic Casimir of the
$(4j+1)$-dimensional representation is vanishing.
Hence the twisted multiplets  $\psi^j_m$ stay with the classical conformal
weight $0$ even at the quantum level.
This result agrees with what is known about twisted chiral primaries by the
algebraic arguments$^{[1,5,20]}$. The bosonic components $u^j_m$ become
physical operators since they are BRST invariant by (25). The BRST partners are
given by $\int dx \rho^j_m$ as usual$^{[5]}$. These physical operators define
the primary states $\vert u^j_m>$ and $\int dx \vert \rho^j_m >$ respectively,
which form an infinite tower of OSp$(2,2)$ multiplets (, precisely speaking,
quantum OSp$(2,2)$ multiplets). Since they have conformal weight $0$,
they are the BRST invariant ground states of the theory. This is reminiscent of
the ground ring of the $c=1$ string$^{[6]}$. But here the ground states are
purely gravitational.
We consider a correlation function of the ground primaries $u^j_m$
$$
<u^{j_1}_{m_1} (x_1) u^{j_2}_{m_2} (x_2) \cdots\cdots
u^{j_N}_{m_N} (x_N)>.
$$
It is just a constant because conformal weight of $u^j_m$
is zero from the quantum group argument. This can be also shown by using the
BRST invariance of $u_j^m$ according to the standard argument of the
topological theory$^{[3]}$. It is worth noting here consistency between the two
assumptions (i) and (ii). An interesting thing with this is that we can
successively exchange the order of the ground primaries $u^j_m$
 by means of the algebra (26). Thus the correlation function is braided by the
$R$-matrices of quantum OSp$(2,2)$.  This braiding property  can not be
extended to correlation functions including the BRST partners $\int dx
\rho^j_m$. The BRST partners  do not obey the exchange algebra (26), although
the fermionic primaries $\rho^j_m$ themselves do well.

\vskip 1cm

In this note we have constructed chiral superconformal fields in the effective
$(2,0)$ supergravity. They are an infinite tower of OSp$(2,2)$ multiplets and
satisfy the classical exchange algebra with the $r$-matrices of the OSp$(2,2)$
group. By the twisting procedure the effective $(2,0)$ supergravity becomes a
topological gravity. From those chiral superconformal fields we have found an
infinite number of BRST invariant quantities with conformal weight $0$ in this
topological gravity, i.e.,
$u^j_m$ and $\int dx \rho^j_m$. These classical arguments are well-based on the
superconformal geometry and the OSp$(2,2)$ group. Hence we were led to assume
naturally that at the quantum level the superconformal symmetry is maintained
and the OSp$(2,2)$ symmetry is took over as the quantum group symmetry. The
first assumption implied that the multiplets $u^j_m$ and $\int dx \rho^j_m$
become BRST invariant primaries at the quantum level. The second assumption was
used to show that
quantization does not modify the classical value of their conformal weights.
Thus we have obtained an infinite number of physical operators with conformal
weight $0$ at the quantum level. They define the BRST invariant ground states
of the theory. Furthermore the second assumption enabled us to discuss that
correlation functions of the multiplets $u^j_m$ are braided by the $R$-matrices
of the quantum OSp$(2,2)$. The last property is not shared by the standard
topological gravity$^{[3,4]}$.

It is still desired to prove  these assumptions rigorously.
In this regard it is interesting to study
 the effective $(2,0)$ supergravity in a conformal gauge, i.e., the $(2,0)$
supersymmetric  Liouville theory. As the $(0,0)$ theory$^{[18,22]}$ it would be
expected to be soluvable.
 There would exist similar quantities to $\psi^j$ and they could be represented
in terms of free fields. It would be then possible to derive our assumptions.

\vskip 2cm

\noindent
{\bf Acknowledgment}

The author thanks the Research Council of K.U. Leuven for the financial
support.

\vfill\eject

\noindent
{\bf References}
\item{1.} E. Witten, Nucl. Phys. B340(1990)281.
\item{2.} K. Li, Nucl. Phys. B354(1991)711;
\item{} T. Eguchi, H. Kanno, Y. Yamada and S.K. Yang, Phys. Lett. B305(1993)
235.
\item{3.} E. Witten, Comm. Math. Phys. 118(1988)411.
\item{4.} J. Labastida, M. Pernici and E. Witten,  B310(1988)611;
\item{} D. Montano and J. Sonnenschein, Nucl. Phys. B313(1989)258;
\item{} J. Distler, Nucl. Phys. B342(1990)523;
\item{} E. Verlinde and H. Verlinde, Nucl. Phys. B348(1991)457.
\item{5.} T. Eguchi and S.K. Yang, Mod. Phys. A5(1990)1693.
\item{6.} A.M. Polyakov, Mod. Phys. Lett. A6(1991)3273;
\item{} D. Gross, I.R. Klebanov and M. Newman, Nucl. Phys. B350(1991)621;
\item{} I.R. Klebanov and A.M. Polyakov, Mod. Phys. Lett. A6(1991)3273;
\item{} E. Witten, Nucl. Phys. B373(1992)187;
\item{} B. Lian and G. Zuckerman, Phys. Lett. B266(1991)21;
\item{} P. Bouwknegt, J. McCarthy and K. Pilch, Comm. Math. Phys. 145(1992)
561.
\item{7.} A.M. Polyakov, Mod. Phys. Lett. A2(1987)843;
\item{} V.G. Knizhnik, A.M. Polyakov and A.B. Zamolodchikov, Mod. Phys. Lett.
A3(1988)819.
\item{8.} A. Alekseev and S. Shatashvili, Nucl. Phys. B323(1989)719.
\item{9.} M. Bershadsky and H. Ooguri, Comm. Math. Phys. 126(1989)49.
\item{10.} A. Neveu and J.-L. Gervais, Nucl. Phys. B209(1982)125;
B238(1984)125.
\item{11.} J. Distler and H. Kawai, Nucl. Phys. B321(1989)509;
\item{} F. David, Mod. Phys. Lett. A3(1988)1651;
\item{} T. Curtright and C. Thorn, Phys. Rev. Lett. 48(1982)1309.
\item{12.} S. Aoyama, Int. J. Mod. Phys. A7(1992)5761.
\item{13.} J.D. Cohn, Nucl. Phys. B284(1987)349;
\item{} P. Di Vecchia, J.L. Petersen and H.B. Zheng, Phys. Lett. B162(1985)327.
\item{14.} H. Aratyn, E. Nissimov, S. Pacheva and S. Solomon, Phys. Lett. B234
(1990)307;
\item{} G. Delius, P. van Nieuwenhuizen and V. Rodgers, Int. J. Mod. Phys.
A5(1990)3943.
\item{15.} S. Aoyama, Mod. Phys. Lett. A6(1991)2069.
\item{16.} A. Alekseev and S. Shatashvili, Comm. Math. Phys. 128(1990)197.
\item{17.} S. Aoyama, Phys. Lett. B256(1991)416.
\item{18.} L.D. Faddeev and L.A. Takhtajan, Springer Lec. Notes in Phys.
246(1986) 166;
\item{} O. Babelon, Phys. Lett. B215(1988)523.
\item{19.} F.A. Smirnov and L.A. Takhtajan, ``Towards a quantum Liouville
theory with $c>1$".
\item{20.} W. Lerche, C. Vafa and N.P. Warner, Nucl. Phys. B324(1989)427.
\item{21.} L. Alvarez-Gaum\'e, C. Gomez and G. Sierra, Phys. Lett.
B220(1989)142.
\item{22.} A. Bilal and J.-L. Gervais, Nucl. Phys. B318(1989) 579;
\item{} J.-L. Gervais, Comm. Math. Phys. 130(1990)257;
\item{} T. Hollowood and P. Mansfield, Nucl. Phys. B330(1990) 720;
\item{} O. Babelon and L. Bonora, Phys. Lett. B253(1991)365;
\item{} V.A. Fateev and S.L. Lukyanov, ``Vertex operators and
representations of quantum universal enveloping algebras",
PAR-LPTHE 91-17.

\bye